\documentclass[aps,prl,twocolumn,groupedaddress,floatfix,showpacs]{revtex4-1}
\usepackage{graphics,graphicx}
\usepackage{color}
\usepackage{amssymb,amsmath}
\usepackage{bm}

\newcommand{\req}[1]{Eq.~(\ref{#1})}
\newcommand{\rref}[1]{(\ref{#1})}

\newcommand{\vare}{\varepsilon}
\newcommand{\Tc}{T_c}

\newcommand{\be}{\begin{equation}}
\newcommand{\ee}{\end{equation}}
\newcommand{\bea}{\begin{eqnarray}}
\newcommand{\eea}{\end{eqnarray}}

\renewcommand{\Re}{{\rm Re}}
\renewcommand{\Im}{{\rm Im}}
\begin{document}
\unitlength = 1mm
\title{Differential conductance of point contacts between an iron-based superconductor and a normal metal}
\author{D. Kuzmanovski}
\author{M. G. Vavilov}
\affiliation{Department of Physics,
             University of Wisconsin, Madison, Wisconsin 53706, USA}

\date{Received 20 May 2011; revised manuscript received 5 August 2011; published 26 August 2011}


\begin{abstract}

We present a theoretical description of the differential conductance of point contacts between a normal metal and a multiband superconductor with extended $s^\pm$ pairing symmetry.  We demonstrate  that the interband impurity scattering broadens the coherent peak near the superconducting gap and significantly reduces its height even at relatively low scattering rates
. This broadening is consistent with a number of recent experiments performed for both tunnel junctions and larger diffusive contacts.   Our theory helps to better evaluate the energy gap of iron-based superconductors from point-contact Andreev spectroscopy measurements.

\end{abstract}
\maketitle

\emph{Introduction.} Iron-pnictide superconductors are multiband materials with several disconnected Fermi surfaces.\cite{Mazin2008,Norman2008,Paglione2010} The structure of the superconducting (SC) energy gap near the Fermi surface (FS) remains one of the unsettled questions of the current investigation of these materials.  Experiments indicate that in some compounds the SC gap vanishes along nodal lines,\cite{Wang2009,Fletcher2009,Hashimoto2010} while many other observations are consistent with a fully gapped FS in SC states.\cite{Chen2008,Gordon2009,Hashimoto2009} But even fully gapped SC states may be realized by different possible structures of the SC order parameter (OP), such as conventional or extended $s^\pm$ symmetry pairings.\cite{Mazin2008,nodes}

Various experimental techniques have been utilized to explore the structure of the SC OP in iron pnictides, including angle-resolved photoemission spectroscopy (ARPES)\cite{ARPES}, NMR\cite{Matano2008} and the London penetration depth technique.\cite{Fletcher2009,Gordon2009,Hashimoto2009,Hashimoto2010,Malone2009, Yong2011} 
More recently, several groups have reported on measurements of the differential conductance of a point contact 
between a single-crystal iron-based pnictide superconductor and a normal metal by a technique called point-contact Andreev spectroscopy (PCAS). Point contacts with both high tunnel barriers\cite{Hanaguri2010,Shan2011,Noat2010} and a large number of conduction channels\cite{Wang2009,Chen2008,Sheet2010,Lu2010,Naidyuk2011} were reported.  In either system  the differential conductance curves were significantly broadened when compared to curves for conventional superconductors within the Blonder-Tinkham-Klapwijk (BTK) theory\cite{Blonder1982} and required the introduction of a somewhat short  phenomenological electron lifetime\cite{Plecenik1994,Noat2010} near the FSs. Previous theoretical analysis\cite{R2} of PCAS for clean multiband  superconductors does not seem to describe the above experiments.

In the present Rapid Communication we argue that the smearing of the differential conductance in PCAS may indicate the realization of  an extended $s^\pm$-wave pairing and moderate interband impurity scattering. For superconductors with extended $s^\pm$-wave pairing, interband impurity scattering mixes quasiparticle states between FSs characterized by opposite signs of the OP 
and has a pair-breaking effect on superconductivity.\cite{Chubukov2008,Vorontsov2009a} A relatively small rate $\Gamma_\pi\ll \Tc$ of the interband scattering (we use $\hbar = 1$ and $k_B=1$) does not significantly change the critical temperature $T_c$ or the SC OPs $\Delta_{e,h}$ on electron and hole pockets of $s^\pm$-wave superconductors. We show, however, that even weak interband scattering significantly  modifies the excitation spectrum near the FSs and reduces a resonance peak in the differential conductance at bias near the SC gap.  

This mechanism of broadening the differential conductance peaks for $s^\pm$-wave superconductors does not involve a phenomenological parameter for the electron life-time. The same model\cite{Vorontsov2009a} was applied to the explanation of the temperature dependence of the London penetration depth.\cite{Gordon2009,Hashimoto2009} The magnetic penetration  depth at low temperatures exhibits a weak nonexponential dependence  on temperature in electron-doped BaFe$_2$As$_2$ materials, consistent with the theoretical temperature dependence for an extended $s^\pm$ SC state in the presence of interband impurity scattering.  Interband impurity scattering also results in a power-law dependence of the spin relaxation rate in nuclear magnetic field resonance.\cite{parker-NMR, Chubukov2008} Therefore, one can evaluate SC OPs  and interband scattering rates from independent experiments.

\emph{Model.}
We consider a simplified two-band model\cite{Chubukov2008} of iron-based pnictide superconductors with a single electron and hole FSs.  Generalization of this model to a  larger number of electron and hole FSs does not qualitatively change our results. We assume that the SC state has an extended $s^\pm$-wave symmetry and is characterized by isotropic OP $\Delta_\alpha$ on electron, $\alpha=e$, or hole, $\alpha=h$, Fermi surfaces, with $\Delta_h\Delta_e<0$. Scattering off disorder can be separated into two categories: intraband scattering when the electron band index stays the same and interband scattering when the electrons change their band index.  We note that the intraband scattering does not affect the SC state with isotropic OP. The interband scattering mixes electron states with opposite OPs in $s^\pm$ wave SC and results in the depairing of the Cooper pairs suppressing superconductivity. In this case, the interband scattering rate $\Gamma_\pi$ characterizes SC properties in a similar way to the spin-scattering rate in conventional superconductors with magnetic impurities.\cite{Maki68}

For a weak enough  current through the point contact, the superconducting state is nearly homogeneous in space.  We disregard the proximity effect in a metallic probe \cite{Volkov1993} and  apply the circuit theory \cite{Nazarov1999} to evaluate the differential conductance $G(V)=dI/dV$:
\be
\label{eq:G}
G(V) = \sum_{\pm, \alpha} \int_{-\infty}^{\infty} \frac{C_{\alpha}(2 T u \pm e V )}{2\cosh^{2}u} d u.
\ee
Here
\be
\begin{split}
\label{eq:C}
C_{\alpha}&(\vare)  =   \Re{(\xi_{\alpha} \, Z_{\alpha}(\xi_{\alpha}))}  \\
 & -  \frac{\Im{Z_{\alpha}(\xi_{\alpha})}}{\Im{\xi_{\alpha}}} \, (\Re{a_{\alpha}})^{2} \left((\Re{\xi_{\alpha}} + 1)^{2} + (\Im{\xi_{\alpha}})^{2}\right) 
\end{split}
\ee
has a meaning of spectral current density and is written in terms of functions $a_\alpha(\vare)$ and $\xi_{\alpha}(\vare) = [1 - a^{2}_{\alpha}(\vare)]/[1 + a^{2}_{\alpha}(\vare)]$ of quasiparticle energy $\vare$. The function $Z_\alpha(x)$ is defined in terms of the transmission eigenvalues $t_n^{(\alpha)}$ of a point contact between states in normal tip and electronic states on Fermi surface $\alpha$:
\be 
\label{eq:Z}
Z_{\alpha}(x)  = \frac{e^{2}}{\pi} \, \sum_{n}{\frac{t^{(\alpha)}_{n}}{2 + t^{(\alpha)}_{n} (x - 1)}}.
\ee

The functions $a_\alpha(\vare)$ are solutions of a system of the following two fourth-order algebraic equations:
\begin{subequations}\label{eq:ricatti}
\begin{eqnarray}
\Delta_{e} (1 - a^{2}_{e}) + 2 i \vare a_{e} & = & \frac{2 \Gamma_{\pi} (a_{e} - a_{h}) (1 + a_{h} a_{e})}{1 + a^{2}_{h}}, \\ 
\Delta_{h} (1 - a^{2}_{h}) + 2 i \vare a_{h} & = & \frac{2 \Gamma_{\pi} (a_{h} - a_{e}) (1 + a_{e} a_{h})}{1 + a^{2}_{e}}.
\end{eqnarray}
\end{subequations}
A proper solution of \req{eq:ricatti} is chosen from the condition that $a_{\alpha}$ behaves as $i \Delta_{\alpha}/[2 (\vare + i \Gamma_{\pi})]$ at $|\vare|\gg |\Delta_{\alpha}|$ and Eqs.~\rref{eq:G}--\rref{eq:Z} recover  the normal conductance $G_N = \left(e^2/2\pi\right) \, \sum_{\alpha,n}t_n^{(\alpha)}$ at bias $eV\gg|\Delta_{e,h}|$. We note that for $\Gamma_\pi=0$, \req{eq:ricatti} describes the BCS-type superconductor and \req{eq:G} recovers the BTK result.  

\begin{figure}[t]
\centerline{\includegraphics[width=\linewidth]{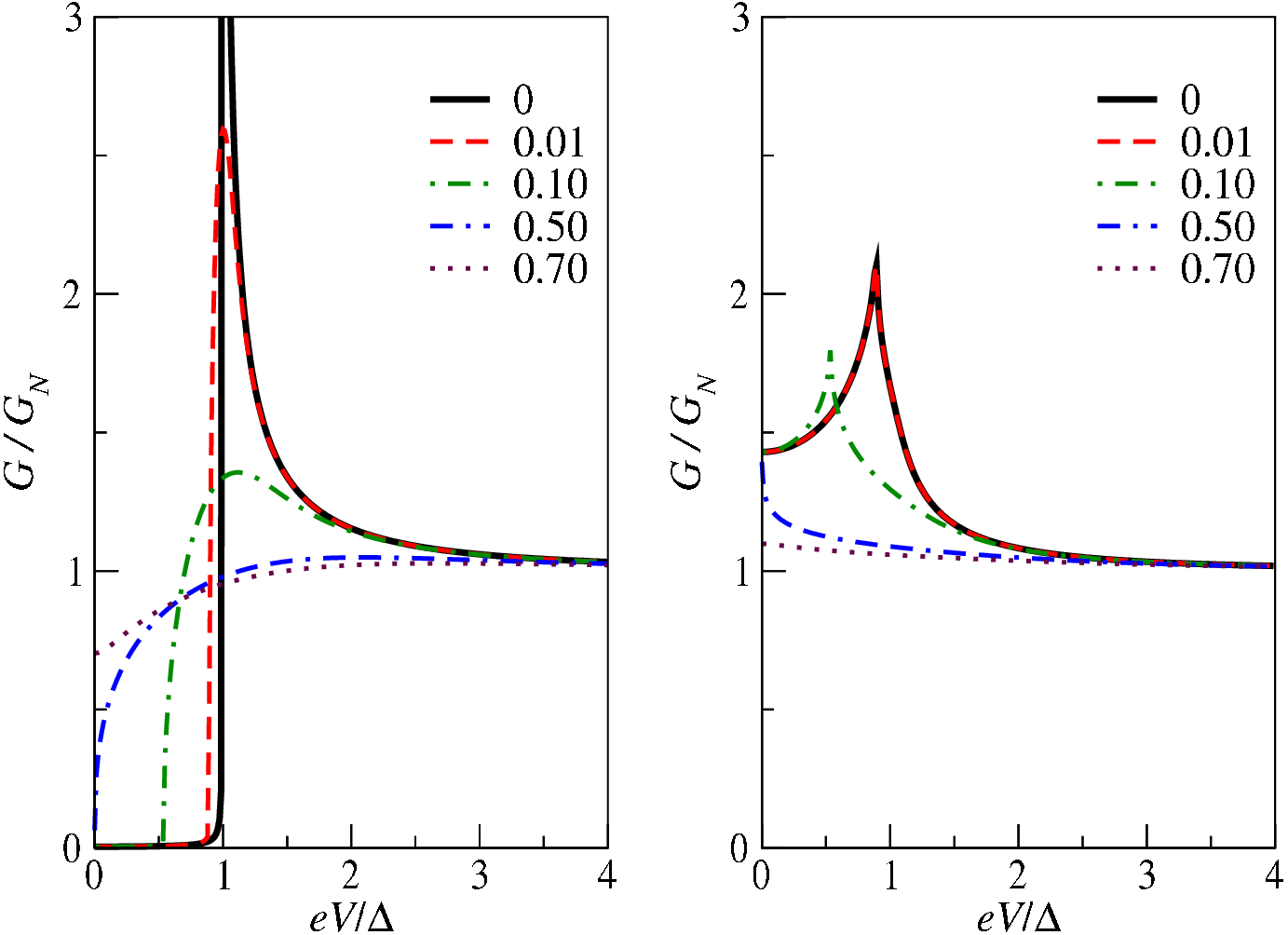}}
\caption{(Color online) Zero-temperature normalized differential conductance curves in the case $\Delta_{h} = -\Delta_{e} = \Delta$ for different values of $\Gamma_{\pi}/\Delta= 0.01,\ 0.1,\ 0.5,\ 0.7$. Left panel: A single--channel tunnel junction with $t^{(h)} = t^{(e)} = 0.01$. Right panel: A diffusive junction with a representative sample of $200$ channels with transmission eigenvalues distributed according to Eq.~(\ref{eq:bimod}) with a mean $\langle t_n^{(h)}\rangle = \langle t_n^{(e)}\rangle = 0.84$.
}
\label{fig:1}
\end{figure}

\emph{Results.}
We first analyze the differential conductance for a superconductor with equal-in-magnitude OPs, 
$|\Delta_{e, h}|=\Delta$. The left panel in Fig.~\ref{fig:1} shows the differential conductance for a single-channel junction with very weak tunnel probability $t^{(\alpha)}\ll 1$ between the normal tip and the superconductor.  

In this case the Andreev reflection is suppressed as $(t_n^{(\alpha)})^2$ and the differential conductance is proportional to the electron density of states (DOS) in a superconductor, except at the very top of Andreev resonance peaks at bias $e \, V=|\Delta_{e,h}|$. As the ratio  $\Gamma_\pi/\Delta$ increases, we observe an evolution of the DOS from a BCS-type DOS with gap equal to $\Delta$ and a very high peak above the gap to a smaller gap and a reduced height of the peak. In particular, a relatively small $\Gamma_\pi/\Delta=0.01$ already drastically reduces the height of the peak.  At $\Gamma_\pi/\Delta\simeq 0.5 $, the gap completely disappears, as expected.\cite{Maki68} We also note that the position of the maximum of the differential conductance moves slowly to higher bias as $\Gamma_\pi/\Delta$ increases.

Larger contacts have many conduction channels with transmission eigenvalues $t_n^{(\alpha)}$ between 0 and 1. Assuming that the contact is diffusive, we model the statistics of $t_n^{(\alpha)}$ by the Dorokhov   distribution:\cite{Beenakker1997}
\be
\label{eq:bimod}
P(t^{(\alpha)}_{n}) = \frac{1}{2 q_{\alpha}t^{(\alpha)}_{n} \, \sqrt{1-t^{(\alpha)}_{n}}}, \quad \mathrm{sech}^{2}{q_{\alpha}} \le t^{(\alpha)}_{n} < 1
\ee
with a relatively high cutoff $q_{\alpha} \sim 1$. The parameter $q_\alpha$ determines the average $\langle t^{(\alpha)}_{n} \rangle = \tanh{q_{\alpha}}/q_{\alpha}$, and the total number of channels can be chosen to match the junction conductance in the normal state $G_N$. We present the differential conductance for such diffusive contacts in case $\Delta_{h} = -\Delta_{e} = \Delta$ for several ratios of $\Gamma_\pi/\Delta$ in the right panel of Fig.~\ref{fig:1}.  
Unlike $G(V)$ of tunnel junctions, $G(V)$ of diffusive contacts may exceed $G_N$ due to the contribution to the current from the Andreev reflection that doubles the conductance of nearly open channels with $t_n^{(\alpha)}\simeq 1$. 
As interband scattering rate $\Gamma_\pi$ increases and suppresses superconductivity, $G(V)$ curves smoothen  and approach $G_N$.  In the gapless regime, $\Gamma_\pi/\Delta\gtrsim 0.5$, the differential conductance exhibits a zero-bias peak. 

\begin{figure}[t]
\centerline{\includegraphics[width=0.7 \linewidth]{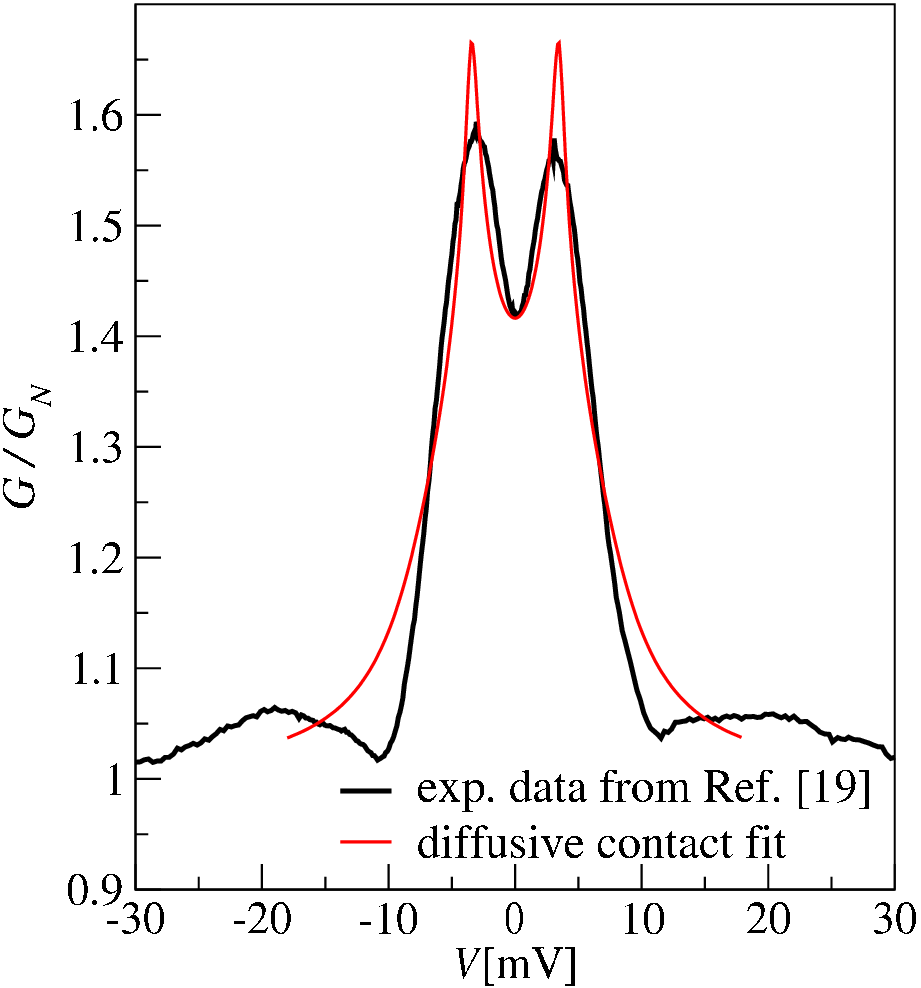}}
\caption{(Color online) A fit to the experimental data of Ref.~\cite[Fig.~3]{Lu2010}, with $\Delta_{h} = -\Delta_{e} = \Delta = 6.3 \, \mathrm{meV}$ and $\Gamma_{\pi} = 0.59 \, \mathrm{meV}$ used as fitting parameters. Only  experimental points for $V>0$ were utilized for the fit. The distribution of transmission eigenvalues, \req{eq:bimod}, was adjusted to match the experimental zero-bias conductance $G(0)$. The temperature was fixed at the reported value  $T=2 \ \mathrm{K}$.
}
\label{fig:2}
\end{figure}

In Fig.~\ref{fig:2} we present a fitting of recent experimental data for diffusive point contacts\cite{Lu2010} using our model of an $s^\pm$ superconductor with equal OPs on the two FSs. Overall, large conductance indicates that there are many channels across the contact between the metallic tip and the $(\mathrm{Ba}_{0.6}\mathrm{K}_{0.4})\mathrm{Fe}_{2}\mathrm{As}_{2}$ superconductor ($T_{c} \sim 37 \, \mathrm{K}$). For an $s^\pm$ superconductor with equal OPs $|\Delta_{e,h}|=\Delta$, the zero-bias conductance $G(0)/G_N$ is independent from $\Delta$ and $\Gamma_\pi$ and depends only on the average value of transmission eigenvalues, chosen as $\langle t_n^{(h)}\rangle = \langle t_n^{(e)}\rangle = 0.84$ to match the data.  Then we found a good fit to experimental data by taking the SC OP $\Delta=|\Delta_{e,h}|=6.3 \, \mathrm{meV}$ and $\Gamma_\pi=0.59 \, \mathrm{meV}$ (see Fig.~\ref{fig:2}). The OP $\Delta$ is somewhat greater than that found in Ref.~\cite{Lu2010} based on a model with a phenomenological electron lifetime.\cite{Plecenik1994} We estimate $2\Delta(0)/T_{c}\approx 3.95$, which is close to the BCS value $3.53$ and consistent with enhancement of $2\Delta/T_c$ due to interband scattering in $s^\pm$ superconductors.\cite{Vorontsov2009a} For the above parameters, we find $\Gamma_\pi/2\pi T_{c, 0} \simeq 0.023$, and the critical temperature  is only moderately suppressed as  $T_c=0.77 \, T_{c,0}$ due to the interband scattering when compared to idealistic $T_{c,0}$ in the absence of scattering. We would like to emphasize that the value of $\Gamma_\pi $ can be estimated for the same sample from independent measurements of magnetic penetration depth or spin relaxation time.

\begin{figure}[t]
\centerline{\includegraphics[width=0.95\linewidth]{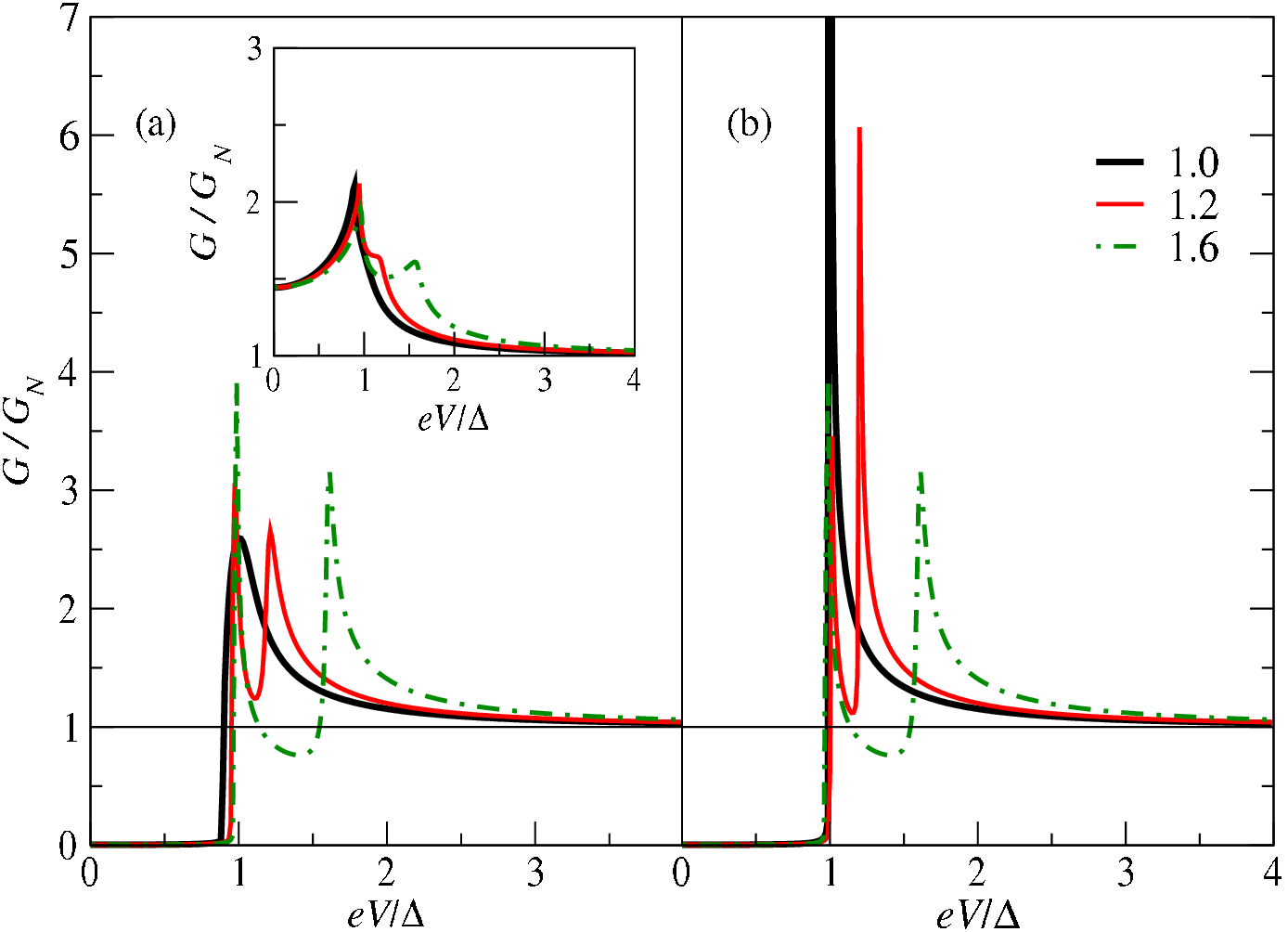}}
\caption{(Color online) Zero-temperature normalized differential conductance curves in the case $\Gamma_{\pi}/\Delta = 0.01$ for different values of the ratio $|\Delta_{e}/\Delta_{h}| = 1.0,\ 1.2,\ 1.6$ ($\Delta_{h} = \Delta$). (a) A single-channel tunnel junction with the sign of the OP opposite on different FSs. (Inset) A diffusive junction with a representative sample of $200$ channels with transmission eigenvalues distributed according to Eq.~(\ref{eq:bimod}), and with a mean $\langle t_n^{(h)}\rangle = \langle t_n^{(e)}\rangle = 0.84$. The sign of the OP is opposite on different FSs. (b) A single-channel tunnel junction with the sign of the OP equal on different FSs.
}
\label{fig:3}
\end{figure}

We also consider the differential conductance for an $s^\pm$ state with different magnitudes of OPs on the FSs, $\Delta_e/\Delta_h\neq -1$. For tunnel junctions with $t^{(\alpha)}\ll 1$, the differential conductance acquires a double-peak feature once the OPs on the two FSs are different [see Fig.~\ref{fig:3}(a)]. We note that once the SC energy gaps  on the FSs have different magnitudes, the resonance peak at smaller bias becomes sharper than that in the case of equal OPs.   
The sharp first peak is a consequence of suppression of the interband scattering of electrons, when one FS is still gapped while the other FS has quasiparticles. This behavior is consistent with the observed differential conductance of tunnel contacts in Ref.~\cite{Shan2011}. As a comparison, Fig.~\ref{fig:3}(b) presents the case of an $s^{++}$ state characterized by equal  signs of the OPs on the two FSs. Although the smearing of the peaks is present even in this case (see also Ref.~\cite{onari2009} for DOS in an $s^{++}$ state), the peaks are much sharper when compared to the $s^{\pm}$ state for  $|\Delta_e/\Delta_h|$ close to unity. When this ratio is   away from unity, the $G(V)$ curves for the $s^\pm$ and $s^{++}$ states become similar, cf. plots for $|\Delta_e/\Delta_h|=1.2$ and  $|\Delta_e/\Delta_h|=1.6$.

In the inset of Fig.~\ref{fig:3}(a) we present the differential conductance of a diffusive point contact. We note that the low-bias coherence peak does not change much when the OPs are taken with different magnitudes, since this peak is already broadened by a simultaneous contribution of the tunneling current above the gap and Andreev reflection within the gap. But the double-peak feature of the plot still remains, and the overall differential conductance peak appears to be broader than that for a model with equal magnitude gaps.   

To investigate temperature dependence of $G(V)$, we consider  $\Delta_h=-\Delta_e=\Delta(T)$.  
In this case the self-consistency  equation for the OP reduces to 
\begin{subequations}
\be
\begin{split}
& \ln\frac{T}{T_c}  +\Psi\left(\frac{1}{2}+\frac{\Gamma_\pi}{\pi T}\right) -\Psi\left(\frac{1}{2}+\frac{\Gamma_\pi}{\pi T_c}\right) \\
& =2\pi T\sum_{\vare_m>0}\left(\frac{y_m}{\Delta(T)\sqrt{y_m^2+1}}-\frac{1}{\vare_m+2\Gamma_\pi}\right),\\
\end{split}
\ee
with an auxiliary variable $y_m$ that satisfies 
\be
[\Delta(T)-y_m \vare_m]\sqrt{y_m^2+1}=2\Gamma_\pi y_m,
\ee
\end{subequations}
and
$\Psi(x)$ being the digamma function. The transition temperature $T_c$ is suppressed by the impurity interband scattering with respect to the clean critical temperature $T_{c,0}$ according to 
$
\ln(T_c/T_{c,0})=\Psi(1/2)-\Psi(1/2+\Gamma_\pi/\pi T_c)
$, and $T_c$ vanishes completely at $\Gamma_\pi\gtrsim 0.07(2 \pi T_{c,0})$. At $\Gamma_\pi \simeq 0.064 (2\pi T_{c,0})$ a SC state becomes gapless and has the critical temperature $T_c\simeq 0.22 \, T_{c,0}$.\cite{Maki68}

\begin{figure}[t]
\centerline{\includegraphics[width=0.96\linewidth]{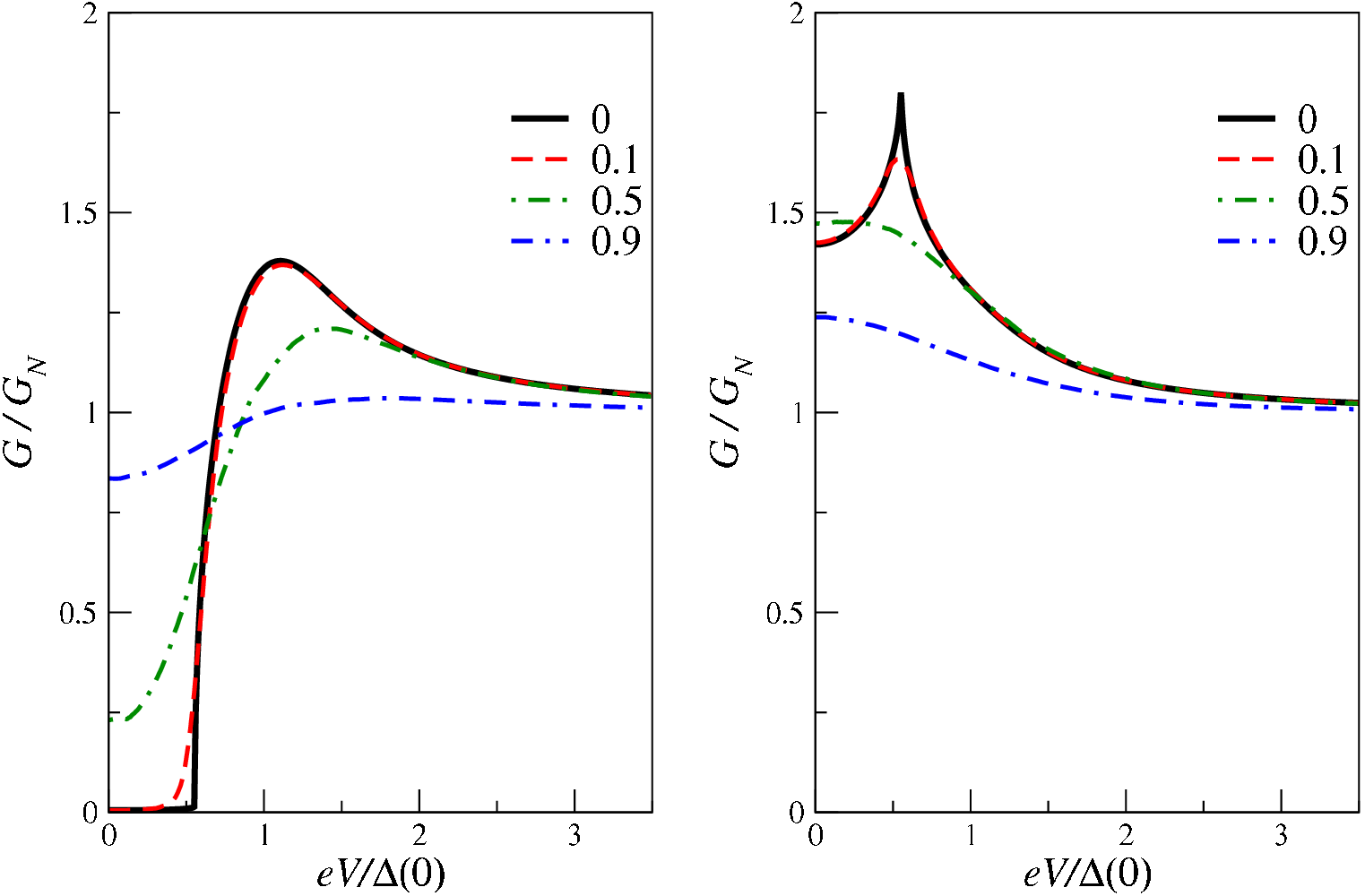}}
\caption{(Color online) Temperature dependence (the varying parameter is $T/T_{c}=0,\ 0.1,\ 0.5,\ 0.9$) of the normalized differential conductance curves in the case $\Delta_{h} = -\Delta_{e} = \Delta(T)$ for a choice of the parameters such that $\Gamma_{\pi}/2\pi T_{c} = 0.029$. Left panel: A single-channel tunnel junction. Right panel: A diffusive junction with a representative sample of $200$ channels with transmission eigenvalues distributed according to Eq.~(\ref{eq:bimod}) and with a mean $\langle t_n^{(h)}\rangle = \langle t_n^{(e)}\rangle = 0.84$.
}
\label{fig:4}
\end{figure}

In Fig.~\ref{fig:4} we plot $G(V)$ for a tunnel junction and a diffusive point contact using $T_{c} = 37 \, \mathrm{K}$ at $\Gamma_{\pi} = 0.59 \, \mathrm{meV}$, found from the fit to experimental data in \cite{Lu2010} with $\Delta(0) = 6.3 \, \mathrm{meV}$. We observe that in tunnel junctions, the positions of the conductance peaks are nearly independent of temperature, but their heights decrease. The value of the zero-bias conductance grows monotonically with temperature until the curve flattens out at the critical temperature.  
In diffusive contact, the coherence peaks are rounded relatively quickly and move to zero bias as temperature increases. At  $T\simeq 0.5 \, T_{c}$ the peak settles at $V=0$, and further increase in temperature leads to the suppression of this zero-bias peak near $T=T_c$.

\emph{In conclusion}, we presented a theoretical description of the differential conductance of a point contact between an iron-based superconductor and a metallic tip. We demonstrated that interband impurity scattering of quasiparticles between Fermi surfaces significantly modifies differential conductance and provides a microscopic depairing mechanism in an $s^\pm$-wave superconducting state. This  behavior stems from the sign-changing superconducting order parameter between Fermi surfaces and is an indirect indicator of $s^\pm$ superconductivity in iron-based superconductors. The  broadening of the differential conductance curves was observed in large contacts.\cite{Sheet2010,Lu2010} Measurements of the differential conductance of tunnel contacts\cite{Hanaguri2010,Shan2011} also showed a significant suppression of the coherence peaks in conductance at  bias near the superconducting gap. These experimental observations 
required an incorporation of a non--vanishing value of the depairing rate, which has a simple microscopic origin in our model.

Electronic band structure and disorder in real materials are likely more complicated than those analyzed here. However, we expect that the relations between thermodynamic and transport characteristics of iron-based superconductors do not crucially depend on details of band structure and disorder and that a single parameter, such as the interband scattering rate~$\Gamma_\pi$, describes them.  Such a parameter can be evaluated from various independent experiments, such as PCAS and  temperature dependence of the magnetic penetration depth that was also fitted in terms of $\Gamma_\pi$.\cite{Vorontsov2009a} Since impurity  scattering rates often depend on conditions of material preparation, a more conclusive analysis would require comparison of results of PCAS and magnetic penetration depth measurements performed on samples produced in similar conditions, as was the case in recent studies of electron doped BaFe$_2$As$_2$ films.\cite{Sheet2010,Yong2011}

We are grateful to V. Chandrasekhar, A. Chubukov, C. Eom, L. Greene, A. Kamenev, A. Levchenko, X. Lu, and W. K. Park for fruitful discussions. We thank the authors of Ref.~\cite{Lu2010} for providing us with their experimental data. This work was partially supported by the Donors of the American Chemical Society Petroleum Research Fund and by NSF Grant DMR No. 0955500.



\begin{thebibliography}{34}
\expandafter\ifx\csname natexlab\endcsname\relax\def\natexlab#1{#1}\fi
\expandafter\ifx\csname bibnamefont\endcsname\relax
  \def\bibnamefont#1{#1}\fi
\expandafter\ifx\csname bibfnamefont\endcsname\relax
  \def\bibfnamefont#1{#1}\fi
\expandafter\ifx\csname citenamefont\endcsname\relax
  \def\citenamefont#1{#1}\fi
\expandafter\ifx\csname url\endcsname\relax
  \def\url#1{\texttt{#1}}\fi
\expandafter\ifx\csname urlprefix\endcsname\relax\def\urlprefix{URL }\fi
\providecommand{\bibinfo}[2]{#2}
\providecommand{\eprint}[2][]{\url{#2}}



\bibitem[{\citenamefont{Mazin et~al.}(2008)\citenamefont{Mazin, Singh,
  Johannes, and Du}}]{Mazin2008}
\bibinfo{author}{\bibfnamefont{I.~I.} \bibnamefont{Mazin}},
  \bibinfo{author}{\bibfnamefont{D.~J.} \bibnamefont{Singh}},
  \bibinfo{author}{\bibfnamefont{M.~D.} \bibnamefont{Johannes}},
  \bibnamefont{and} \bibinfo{author}{\bibfnamefont{M.~H.} \bibnamefont{Du}},
  \bibinfo{journal}{Phys. Rev. Lett.} \textbf{\bibinfo{volume}{101}},
  \bibinfo{pages}{057003} (\bibinfo{year}{2008}).

\bibitem[{\citenamefont{Norman}({2008})}]{Norman2008}
\bibinfo{author}{\bibfnamefont{M.~R.} \bibnamefont{Norman}},
  \bibinfo{journal}{{Physics}} \textbf{\bibinfo{volume}{{1}}},
  \bibinfo{pages}{21} (\bibinfo{year}{{2008}}).

\bibitem[{\citenamefont{Paglione and Greene}(2010)}]{Paglione2010}
\bibinfo{author}{\bibfnamefont{J.}~\bibnamefont{Paglione}} \bibnamefont{and}
  \bibinfo{author}{\bibfnamefont{R.~L.} \bibnamefont{Greene}},
  \bibinfo{journal}{Nat. Phys.} \textbf{\bibinfo{volume}{6}},
  \bibinfo{pages}{645} (\bibinfo{year}{2010}).

\bibitem[{\citenamefont{Wang et~al.}(2009)\citenamefont{Wang, Shan, Fang,
  Cheng, Ren, and Wen}}]{Wang2009}
\bibinfo{author}{\bibfnamefont{Y.-L.} \bibnamefont{Wang}},
  \bibinfo{author}{\bibfnamefont{L.}~\bibnamefont{Shan}},
\bibnamefont{et~al.},
  \bibinfo{journal}{Supercond. Sci. Technol.} \textbf{\bibinfo{volume}{22}},
  \bibinfo{pages}{015018} (\bibinfo{year}{2009}).

\bibitem[{\citenamefont{Fletcher et~al.}(2009)\citenamefont{Fletcher, Serafin,
  Malone, Analytis, Chu, Erickson, Fisher, and Carrington}}]{Fletcher2009}
\bibinfo{author}{\bibfnamefont{J.~D.} \bibnamefont{Fletcher}},
  \bibinfo{author}{\bibfnamefont{A.}~\bibnamefont{Serafin}},
\bibnamefont{et~al.},
  \bibinfo{journal}{Phys. Rev. Lett.} \textbf{\bibinfo{volume}{102}},
  \bibinfo{pages}{147001} (\bibinfo{year}{2009}).

\bibitem[{\citenamefont{Hashimoto et~al.}(2010)\citenamefont{Hashimoto,
  Yamashita, Kasahara, Senshu, Nakata, Tonegawa, Ikada, Serafin, Carrington,
  Terashima et~al.}}]{Hashimoto2010}
\bibinfo{author}{\bibfnamefont{K.}~\bibnamefont{Hashimoto}},
  \bibinfo{author}{\bibfnamefont{M.}~\bibnamefont{Yamashita}},
  \bibnamefont{et~al.}, \bibinfo{journal}{Phys. Rev. B}
  \textbf{\bibinfo{volume}{81}}, \bibinfo{pages}{220501}
  (\bibinfo{year}{2010}).

\bibitem[{\citenamefont{Chen et~al.}(2008)\citenamefont{Chen, Tesanovic, Liu,
  Chen, and Chien}}]{Chen2008}
\bibinfo{author}{\bibfnamefont{T.~Y.} \bibnamefont{Chen}},
  \bibinfo{author}{\bibfnamefont{Z.}~\bibnamefont{Tesanovic}},
\bibnamefont{et~al.},
  \bibinfo{journal}{Nature} \textbf{\bibinfo{volume}{453}},
  \bibinfo{pages}{1224} (\bibinfo{year}{2008}).

\bibitem[{\citenamefont{Gordon et~al.}(2009)\citenamefont{Gordon, Martin, Kim,
  Ni, Tanatar, Schmalian, Mazin, Bud’ko, Canfield, and
  Prozorov}}]{Gordon2009}
\bibinfo{author}{\bibfnamefont{R.}~\bibnamefont{Gordon}},
  \bibinfo{author}{\bibfnamefont{C.}~\bibnamefont{Martin}},
\bibnamefont{et~al.},
  \bibinfo{journal}{Phys. Rev. B} \textbf{\bibinfo{volume}{79}},
  \bibinfo{pages}{100506} (\bibinfo{year}{2009}).

\bibitem[{\citenamefont{{Hashimoto} et~al.}(2009)\citenamefont{{Hashimoto},
  {Shibauchi}, {Kasahara}, {Ikada}, {Tonegawa}, {Kato}, {Okazaki}, {van der
  Beek}, {Konczykowski}, {Takeya} et~al.}}]{Hashimoto2009}
\bibinfo{author}{\bibfnamefont{K.}~\bibnamefont{{Hashimoto}}},
  \bibinfo{author}{\bibfnamefont{T.}~\bibnamefont{{Shibauchi}}},
  \bibnamefont{et~al.}, \bibinfo{journal}{Phys. Rev. Lett.}
  \textbf{\bibinfo{volume}{102}}, \bibinfo{pages}{207001}
  (\bibinfo{year}{2009}).

\bibitem{nodes}
\bibinfo{author}{\bibfnamefont{T.~A.} \bibnamefont{Maier}},
  \bibinfo{author}{\bibfnamefont{S.}~\bibnamefont{Graser}},
  \bibinfo{author}{\bibfnamefont{D.~J.} \bibnamefont{Scalapino}},
  \bibnamefont{and} \bibinfo{author}{\bibfnamefont{P.~J.}
  \bibnamefont{Hirschfeld}}, \bibinfo{journal}{Phys. Rev. B}
  \textbf{\bibinfo{volume}{79}}, \bibinfo{pages}{224510}
  (\bibinfo{year}{2009});
\bibinfo{author}{\bibfnamefont{A.~V.} \bibnamefont{Chubukov}},
  \bibinfo{author}{\bibfnamefont{M.~G.} \bibnamefont{Vavilov}},
  \bibnamefont{and} \bibinfo{author}{\bibfnamefont{A.~B.}
  \bibnamefont{Vorontsov}}, \bibinfo{journal}{Phys. Rev. B}
  \textbf{\bibinfo{volume}{80}}, \bibinfo{pages}{140515}
  (\bibinfo{year}{2009});
\bibinfo{author}{\bibfnamefont{R.}~\bibnamefont{Thomale}},
  \bibinfo{author}{\bibfnamefont{C.}~\bibnamefont{Platt}},
  \bibinfo{author}{\bibfnamefont{J.}~\bibnamefont{Hu}},
  \bibinfo{author}{\bibfnamefont{C.}~\bibnamefont{Honerkamp}},
  \bibnamefont{and} \bibinfo{author}{\bibfnamefont{B.~A.}
  \bibnamefont{Bernevig}}, \bibinfo{journal}{Phys. Rev. B}
  \textbf{\bibinfo{volume}{80}}, \bibinfo{pages}{180505}
  (\bibinfo{year}{2009}).

\bibitem[{\citenamefont{Nakayama et~al.}(2011)\citenamefont{Nakayama, Sato,
  Richard, Xu, Kawahara, Umezawa, Qian, Neupane, Chen, Ding
  et~al.}}]{ARPES}
\bibinfo{author}{\bibfnamefont{K.}~\bibnamefont{Nakayama}},
  \bibinfo{author}{\bibfnamefont{T.}~\bibnamefont{Sato}},
\bibnamefont{et~al.},
  \bibinfo{journal}{Phys. Rev. B} \textbf{\bibinfo{volume}{83}},
  \bibinfo{pages}{020501} (\bibinfo{year}{2011});
\bibinfo{author}{\bibfnamefont{X.-P.} \bibnamefont{Wang}},
  \bibinfo{author}{\bibfnamefont{T.}~\bibnamefont{Qian}},
\bibnamefont{et~al.},
  \bibinfo{journal}{Europhys. Lett.} \textbf{\bibinfo{volume}{93}},
  \bibinfo{pages}{57001} (\bibinfo{year}{2011}).

\bibitem[{\citenamefont{Matano et~al.}(2008)\citenamefont{Matano, Ren, Dong,
  Sun, Zhao, and Zheng}}]{Matano2008}
\bibinfo{author}{\bibfnamefont{K.}~\bibnamefont{Matano}},
  \bibinfo{author}{\bibfnamefont{Z.~A.} \bibnamefont{Ren}},
\bibnamefont{et~al.},
  \bibinfo{journal}{Europhys. Lett.} \textbf{\bibinfo{volume}{83}},
  \bibinfo{pages}{57001} (\bibinfo{year}{2008}).

\bibitem[{\citenamefont{Malone et~al.}(2009)\citenamefont{Malone, Fletcher,
  Serafin, and Carrington}}]{Malone2009}
\bibinfo{author}{\bibfnamefont{L.}~\bibnamefont{Malone}},
  \bibinfo{author}{\bibfnamefont{J.~D.} \bibnamefont{Fletcher}}, \bibnamefont{et~al.},
  \bibinfo{journal}{Phys. Rev. B} \textbf{\bibinfo{volume}{79}},
  \bibinfo{pages}{140501} (\bibinfo{year}{2009}).

\bibitem[{\citenamefont{Yong et~al.}(2011)\citenamefont{Yong, Lee, Jiang, Bark,
  Weiss, Hellstrom, Larbalestier, Eom, and Lemberger}}]{Yong2011}
\bibinfo{author}{\bibfnamefont{J.}~\bibnamefont{Yong}},
  \bibinfo{author}{\bibfnamefont{S.}~\bibnamefont{Lee}},
\bibnamefont{et~al.},
  \bibinfo{journal}{Phys. Rev. B} \textbf{\bibinfo{volume}{83}},
  \bibinfo{pages}{104510} (\bibinfo{year}{2011}).
  
\bibitem[{\citenamefont{Hanaguri et~al.}(2010)\citenamefont{Hanaguri, Niitaka,
  Kuroki, and Takagi}}]{Hanaguri2010}
\bibinfo{author}{\bibfnamefont{T.}~\bibnamefont{Hanaguri}},
  \bibinfo{author}{\bibfnamefont{S.}~\bibnamefont{Niitaka}},
  \bibinfo{author}{\bibfnamefont{K.}~\bibnamefont{Kuroki}}, \bibnamefont{and}
  \bibinfo{author}{\bibfnamefont{H.}~\bibnamefont{Takagi}},
  \bibinfo{journal}{Science} \textbf{\bibinfo{volume}{328}},
  \bibinfo{pages}{474} (\bibinfo{year}{2010}).

\bibitem[{\citenamefont{Shan et~al.}(2011)\citenamefont{Shan, Wang, Gong, Shen,
  Huang, Yang, Ren, and Wen}}]{Shan2011}
\bibinfo{author}{\bibfnamefont{L.}~\bibnamefont{Shan}},
  \bibinfo{author}{\bibfnamefont{Y.-L.} \bibnamefont{Wang}},
\bibnamefont{et~al.},
  \bibinfo{journal}{Phys. Rev. B} \textbf{\bibinfo{volume}{83}},
  \bibinfo{pages}{060510} (\bibinfo{year}{2011}).
  
  
\bibitem{Noat2010} Y.~Noat, T.~Cren, \bibnamefont{et~al.}, \bibinfo{journal}{J. Phys.: Condens. Matter} \textbf{\bibinfo{volume}{22}},
  \bibinfo{pages}{465701} (\bibinfo{year}{2010}). 

\bibitem[{\citenamefont{Sheet et~al.}(2010)\citenamefont{Sheet, Mehta, Dikin,
  Lee, Bark, Jiang, Weiss, Hellstrom, Rzchowski, Eom et~al.}}]{Sheet2010}
\bibinfo{author}{\bibfnamefont{G.}~\bibnamefont{Sheet}},
  \bibinfo{author}{\bibfnamefont{M.}~\bibnamefont{Mehta}},
\bibnamefont{et~al.},
  \bibinfo{journal}{Phys. Rev. Lett.} \textbf{\bibinfo{volume}{105}},
  \bibinfo{pages}{167003} (\bibinfo{year}{2010}).

\bibitem[{\citenamefont{Lu et~al.}(2010)\citenamefont{Lu, Park, Yuan, Chen,
  Luo, Wang, Sefat, McGuire, Jin, Sales et~al.}}]{Lu2010}
\bibinfo{author}{\bibfnamefont{X.}~\bibnamefont{Lu}},
  \bibinfo{author}{\bibfnamefont{W.~K.} \bibnamefont{Park}},
  \bibnamefont{et~al.}, \bibinfo{journal}{Supercond. Sci. Technol.}
  \textbf{\bibinfo{volume}{23}}, \bibinfo{pages}{054009}
  (\bibinfo{year}{2010}).
  
\bibitem{Naidyuk2011}
\bibinfo{author}{\bibfnamefont{Yu.~G.}~\bibnamefont{Naidyuk}},
  \bibinfo{author}{\bibfnamefont{O.~E.}~\bibnamefont{Kvitnitskaya}},
  \bibnamefont{et~al.}, \bibinfo{journal}{Supercond. Sci. Technol.}
  \textbf{\bibinfo{volume}{24}}, \bibinfo{pages}{065010}
  (\bibinfo{year}{2011}).

\bibitem[{\citenamefont{Blonder et~al.}(1982)\citenamefont{Blonder, Tinkham,
  and Klapwijk}}]{Blonder1982}
\bibinfo{author}{\bibfnamefont{G.~E.}~\bibnamefont{Blonder}},
  \bibinfo{author}{\bibfnamefont{M.}~\bibnamefont{Tinkham}}, \bibnamefont{and}
  \bibinfo{author}{\bibfnamefont{T.~M.}~\bibnamefont{Klapwijk}},
  \bibinfo{journal}{Phys. Rev. B} \textbf{\bibinfo{volume}{25}},
  \bibinfo{pages}{4515} (\bibinfo{year}{1982}).

\bibitem[{\citenamefont{Plecen\'{\i}k et~al.}(1994)\citenamefont{Plecen\'{\i}k,
  Grajcar, and Beňa\v{c}ka}}]{Plecenik1994}
\bibinfo{author}{\bibfnamefont{A.}~\bibnamefont{Plecen\'{\i}k}},
  \bibinfo{author}{\bibfnamefont{M.}~\bibnamefont{Grajcar}}, \bibnamefont{et~al.},
  \bibinfo{journal}{Phys. Rev. B} \textbf{\bibinfo{volume}{49}},
  \bibinfo{pages}{10016} (\bibinfo{year}{1994}).

\bibitem[{\citenamefont{Linder and Sudb\o{}}(2009)}]{R2}
\bibinfo{author}{\bibfnamefont{J.}~\bibnamefont{Linder}} \bibnamefont{and}
  \bibinfo{author}{\bibfnamefont{A.}~\bibnamefont{Sudb\o{}}},
  \bibinfo{journal}{Phys. Rev. B} \textbf{\bibinfo{volume}{79}},
  \bibinfo{pages}{020501} (\bibinfo{year}{2009}).
  
\bibitem[{\citenamefont{Chubukov et~al.}(2008)\citenamefont{Chubukov, Efremov,
  and Eremin}}]{Chubukov2008}
\bibinfo{author}{\bibfnamefont{A.~V.} \bibnamefont{Chubukov}},
  \bibinfo{author}{\bibfnamefont{D.~V.}~\bibnamefont{Efremov}}, \bibnamefont{and}
  \bibinfo{author}{\bibfnamefont{I.}~\bibnamefont{Eremin}},
  \bibinfo{journal}{Phys. Rev. B} \textbf{\bibinfo{volume}{78}},
  \bibinfo{pages}{134512} (\bibinfo{year}{2008}).

\bibitem[{\citenamefont{Vorontsov et~al.}(2009)\citenamefont{Vorontsov,
  Vavilov, and Chubukov}}]{Vorontsov2009a}
\bibinfo{author}{\bibfnamefont{A.~B.} \bibnamefont{Vorontsov}},
  \bibinfo{author}{\bibfnamefont{M.~G.} \bibnamefont{Vavilov}},
  \bibnamefont{and} \bibinfo{author}{\bibfnamefont{A.~V.}
  \bibnamefont{Chubukov}}, \bibinfo{journal}{Phys. Rev. B}
  \textbf{\bibinfo{volume}{79}}, \bibinfo{pages}{140507}
  (\bibinfo{year}{2009}).

\bibitem[{\citenamefont{Parker et~al.}(2008)\citenamefont{Parker, Dolgov,
  Korshunov, Golubov, and Mazin}}]{parker-NMR}
\bibinfo{author}{\bibfnamefont{D.}~\bibnamefont{Parker}},
  \bibinfo{author}{\bibfnamefont{O.~V.} \bibnamefont{Dolgov}},
  \bibinfo{author}{\bibfnamefont{M.~M.} \bibnamefont{Korshunov}},
  \bibinfo{author}{\bibfnamefont{A.~A.} \bibnamefont{Golubov}},
  \bibnamefont{and} \bibinfo{author}{\bibfnamefont{I.~I.} \bibnamefont{Mazin}},
  \bibinfo{journal}{Phys. Rev. B} \textbf{\bibinfo{volume}{78}},
  \bibinfo{pages}{134524} (\bibinfo{year}{2008}).

\bibitem{Maki68}
\bibinfo{author}\bibnamefont{A.}~\bibnamefont{A.}~\bibnamefont{Abrikosov} \bibnamefont{and} \bibnamefont{L.}~\bibnamefont{P.}~\bibnamefont{Gor'kov},
  \bibinfo{journal}{Zh. Eksp. Teor. Fiz. \textbf{\bibinfo{volume}{39}}},
  \bibinfo{pages}{1781} (\bibinfo{year}{1961})
  [\bibinfo{journal}{Sov. Phys. JETP} \textbf{\bibinfo{volume}{12}},
      \bibinfo{pages}{1243}
      (\bibinfo{year}{1961})];
\bibinfo{author}{\bibfnamefont{K.}~\bibnamefont{Maki}},
  \bibinfo{journal}{{Prog. Theor. Phys.}} \textbf{\bibinfo{volume}{{39}}},
  \bibinfo{pages}{897} (\bibinfo{year}{{1968}}).

\bibitem[{\citenamefont{Volkov et~al.}(1993)\citenamefont{Volkov, Zaitsev, and
  Klapwijk}}]{Volkov1993}
\bibinfo{author}{\bibfnamefont{A.}~\bibnamefont{Volkov}},
  \bibinfo{author}{\bibfnamefont{A.}~\bibnamefont{Zaitsev}}, \bibnamefont{and}
  \bibinfo{author}{\bibfnamefont{T.}~\bibnamefont{Klapwijk}},
  \bibinfo{journal}{Physica C} \textbf{\bibinfo{volume}{210}},
  \bibinfo{pages}{21 } (\bibinfo{year}{1993}).

\bibitem[{\citenamefont{Nazarov}(1999)}]{Nazarov1999}
\bibinfo{author}{\bibfnamefont{Y.~V.} \bibnamefont{Nazarov}},
  \bibinfo{journal}{Superlattices Microstruct.} \textbf{\bibinfo{volume}{25}},
  \bibinfo{pages}{1221} (\bibinfo{year}{1999}).

\bibitem[{\citenamefont{Beenakker}(1997)}]{Beenakker1997}
\bibinfo{author}{\bibfnamefont{C.}~\bibnamefont{Beenakker}},
  \bibinfo{journal}{Rev. Mod. Phys.} \textbf{\bibinfo{volume}{69}},
  \bibinfo{pages}{731} (\bibinfo{year}{1997}).

\bibitem[{\citenamefont{Onari and Kontani}(2009)}]{onari2009}
\bibinfo{author}{\bibfnamefont{S.}~\bibnamefont{Onari}} \bibnamefont{and}
  \bibinfo{author}{\bibfnamefont{H.}~\bibnamefont{Kontani}},
  \bibinfo{journal}{Phys. Rev. Lett.} \textbf{\bibinfo{volume}{103}},
  \bibinfo{pages}{177001} (\bibinfo{year}{2009}).

\end{thebibliography}

\end{document}